\documentclass[runningheads]{llncs}
\usepackage{amsmath,amssymb,amsfonts}
\usepackage{marvosym}
\usepackage[T1]{fontenc}
\usepackage{multirow}
\usepackage{cite, hyperref, tikz}
\hypersetup{
            colorlinks=true,
            linkcolor=blue,
            anchorcolor=blue,
            citecolor=blue
            }

\usepackage{graphicx}
\begin{document}

\title{Pubic Symphysis-Fetal Head Segmentation Network Using BiFormer Attention Mechanism and Multipath Dilated Convolution}

%

\author{Pengzhou Cai$^{\ast}$ \and
Lu Jiang$^{\ast}$ \and
Yanxin Li$^{\ast}$ \and Xiaojuan Liu$^{\dagger}$ \and Libin Lan$^{\ast}$\textsuperscript{(\Letter)}}

\authorrunning{P. Cai et al.}

\institute{$^{\ast}$College of Computer Science and Engineering, Chongqing University of Technology, Chongqing, China \\
$^{\dagger}$ College of Artificial Intelligence, Chongqing University of Technology, Chongqing, China \\
\email{lanlbn@cqut.edu.cn}\\
}

\maketitle              

\begin{abstract}

Pubic symphysis-fetal head segmentation in transperineal ultrasound images plays a critical role for the assessment of fetal head descent and progression. Existing transformer segmentation methods based on sparse attention mechanism use handcrafted static patterns, which leads to great differences in terms of segmentation performance on specific datasets. To address this issue, we introduce a dynamic, query-aware sparse attention mechanism for ultrasound image segmentation. Specifically, we propose a novel method, named BRAU-Net to solve the pubic symphysis-fetal head segmentation task in this paper. The method adopts a U-Net-like encoder-decoder architecture with bi-level routing attention and skip connections, which effectively learns local-global semantic information. In addition, we propose an inverted bottleneck patch expanding (IBPE) module to reduce information loss while performing up-sampling operations. The proposed BRAU-Net is evaluated on FH-PS-AoP and HC18 datasets. The results demonstrate that our method could achieve excellent segmentation results. The code is available on \href{https://github.com/Caipengzhou/BRAU-Net}{GitHub}.

\end{abstract}

\section{Introduction}
\begin{figure}[htbp]
\centering
\includegraphics[width=0.8\linewidth, keepaspectratio]{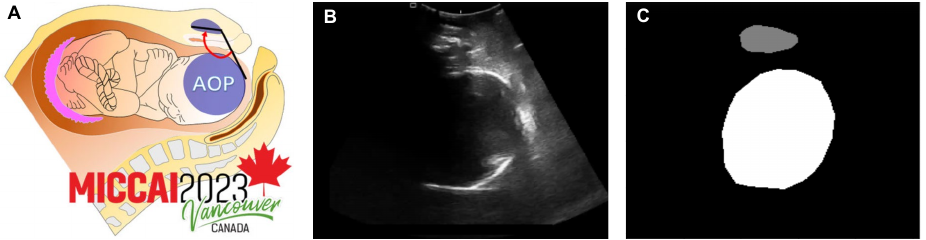}
\caption{(A) The fetal head-pubic symphysis segmentation challenge. (B) A transperineal ultrasound image exhibits the pubic symphysis (PS) and fetal head (FH). (C) The mask of the PS (in grey) and FH (in white).}
\label{fig1:t2tattention}
\end{figure}

A key factor in the increased risk of maternal and perinatal morbidity is prolonged labor due to delayed fetal descent \cite{RN1}. However, accurately locating the fetal head during descent remains a major challenge \cite{RN2}. To address the demand for a more objective diagnosis, transperineal ultrasound (TPU) has emerged as a viable solution. Manual segmentation of the pubis symphysis (PS) and fetal head (FH) from intrauterine (ITU) images is currently considered as the most dependable method (see Fig. \ref{fig1:t2tattention} for a segmentation case), but it is exceptionally time-consuming and susceptible to subjectivity \cite{RN4}. Therefore, an accurate automatic segmentation algorithm for transperineal ultrasound images analysis seems to be crucial for reducing the risk of maternal and perinatal morbidity.

With the rapid development of convolutional neural networks (CNNs), fully convolutional network (FCN) \cite{fcn} generated pixel-wise predictions for segmentation tasks by replacing full connected layers with convolutional layers. The U-Net \cite{unet} designed skip connections in each scale of the  model to enhance the FCN architecture. Afterward, many variants of U-Net have been proposed to perform 2D and 3D medical image segmentation tasks and made outstanding success in a wide range of medical applications. Attention U-Net \cite{attentionunet} integrated attention gates (AGs) to capture complex spatial dependencies more efficiently and improved overall performance. 
MT-Unet \cite{muititask} used ResBlock to build a decoder for segmenting pubic symphysis and fetal heads from TPU images. 
DBSN \cite{DBSN} adopted lower and upper double branches as decoder component of U-shaped architecture to improve generalization performance.
However, these CNN-based methods generally demonstrate limitations in explicitly modeling long-range dependency on account of the inherent locality of convolution operations.

In order to overcome the limitations of CNNs, researchers have tried to introduce Transformer \cite{transformer} into CNN to improve the performance of the network. TransUNet \cite{transunet} was a hybrid CNN-Transformer U-Net-like architecture designed for medical image segmentation, combining the benefits of CNN and Transformer. It leveraged self-attention mechanism to capture global contextual information while maintaining spatial details critical for accurate segmentation. Similar to TransUNet, Transfuse \cite{transfuse} combined Transformer and CNN in parallel to improve the segmentation capability of the model. Additionally, HiFormer \cite{hiformer} and contextual attention network \cite{transformerrmeetunet} effectively bridged CNN and Transformer for medical image segmentation. Nevertheless, their self-attention mechanisms exhibit quadratic computational complexity along with the number of tokens.

In order to reduce the quadratic computational complexity of the transformer models, some works have introduced sparse transformer into computer vision. In the medical segmentation domain, SwinUNet \cite{swinunet} introduced a groundbreaking synergy between the Swin Transformer \cite{swintransformer} and U-Net architecture. It employed two consecutive transformer blocks with different window settings to recapture contextual semantic features from neighboring windows, aiming to revolutionize performance in medical image segmentation. 
MedT \cite{medt} proposed a gated position-sensitive axial attention mechanism to extend the existing transformer architectures, which makes it possible to effectively train models on small-scale medical datasets.
However, the sparse transformers used by these methods are manually designed and static, which leads to large differences on segmentation performance among specific datasets. 

In order to address the above limitations, we introduce a dynamic, query-aware sparse attention mechanism for transperineal ultrasound image segmentation. For this purpose, we attempt to adopt BiFormer block \cite{biformer}, the core part of which is bi-level routing attention,  as a basic unit to build the encoder-decoder architecture for local-global semantic information learning. Furthermore, to reduce the information loss caused by dimensional compression and obtain a better spatial contextual semantic information, an inverted bottleneck patch expanding module, short as IBPE, is proposed and integrated in the network. Our result in the PSFHS challenge \footnote{https://ps-fh-aop-2023.grand-challenge.org/ \label{synapse}} verifies that our model has excellent segmentation performance on FH-PS-AoP dataset. It is worth noting that only a variant of BRAU-Net without pre-training weights initialization and a patch expanding layer \cite{swinunet} as up-sampling were used in the PSFHS challenge, and our result won the 7-$th$ place on the segmentation task. 

As the extended version of BRAU-Net, the main contributions of our work are summarized as follows:  \romannumeral 1) We propose a novel network based on U-shape encoder-decoder architecture, named BRAU-Net which uses bi-level routing attention as core building block for tackle the pubic symphysis-fetal head segmentation, therefore it can effectively learn local-global semantic information while reducing computational complexity. \romannumeral 2) We design an inverted bottleneck patch expanding module, short as IBPE, which can reduce information loss caused by dimensional compression and gain better spatial contextual semantic information. \romannumeral 3) We evaluate the proposed BRAU-Net on FH-PS-AoP and HC18 datasets. The results demonstrate that the proposed BRAU-Net has achieved excellent results on both datasets.

\section{Related Work}
\subsection{Ultrasonic Image Segmentation}
 Convolutional neural networks (CNNs) have made significant breakthroughs in the past ten years. The U-Net \cite{unet} was firstly proposed for cell segmentation and demonstrated remarkable performance. Researchers have developed many excellent models for ultrasonic image segmentation based on U-shaped networks. NU-Net \cite{nunet} improved both the segmentation accuracy of breast tumors and the ability to characterize target or regional features by reducing. SMUNet \cite{smunet} consisted of a main network attached to the midstream and an auxiliary network, which was a significantly guided morphological sensing U-Net for focal segmentation in breast ultrasound (BUS) images. 
 The FH-PSSNet \cite{fh-pssnet} incorporated a dual attention module, a multi-scale feature screening module and a direction guidance block to an encoder–decoder U-shaped network used to automatic angle of progression (AoP) measurement. DBSN \cite{DBSN} used the feature maps from lower branch as the input of the upper branch to assist in decoding, and exploited attention gates (AGs) to provide advanced semantic information to refine the segmentation in the upper path.
 In \cite{muititask}, a multi-task convolutional neural network was proposed to automatically measure AoP values, where for segmentation tasks, channel attention was introduced in the shared feature encoder to improve the robustness of layer encoding. The above methods are all segmentation methods based on convolutional neural networks and show excellent generalization ability and robustness in ultrasonic image segmentation. In order to solve the limitations of CNN in long-term dependence modeling, we aim to find new solutions for ultrasonic image segmentation from the perspective of dynamic sparse transformer.
 \subsection{Upsampling Methods}
 Upsampling is a common technique used in image processing to increase the resolution of images. Upsampling methods can be divided into linear interpolation methods and deep learning based methods. The former mainly includes some traditional methods such as nearest neighbor interpolation and bilinear interpolation, and bicubic interpolation. However, these methods have limitations in capturing complex relationships between pixels, and thus some semantic information may be lost during interpolation operations. This will lead to blurred image edges and even poor image quality. As for the latter, transposed convolution and interpolated convolution have been used to implement upsampling operations. Transposed convolution can cause the heavy computation, and the interpolation convolution is easy to bring about information. In SwinUNet \cite{swinunet}, the empirical results demonstrated that using the expanding layer as up-sampling method is more applicable than the above traditional up-sampling ones, but it has a disadvantage that may incur information loss due to dimensional compression.
 In summary, the above limitations prompted us to design the IBPE module that makes up for the shortcomings of the expanding layer.
 
\section{Method}
\label{sec:format}
In this section, we start by describing the overall architecture of the proposed BRAU-Net. We then introduce the BiFormer block. Finally, we specify the inverted bottleneck patch expanding (IBPE) module.
\subsection{BRAU-Net Architecture}
\begin{figure*}[htbp]
\centering
\includegraphics[width=1.0\linewidth, keepaspectratio]{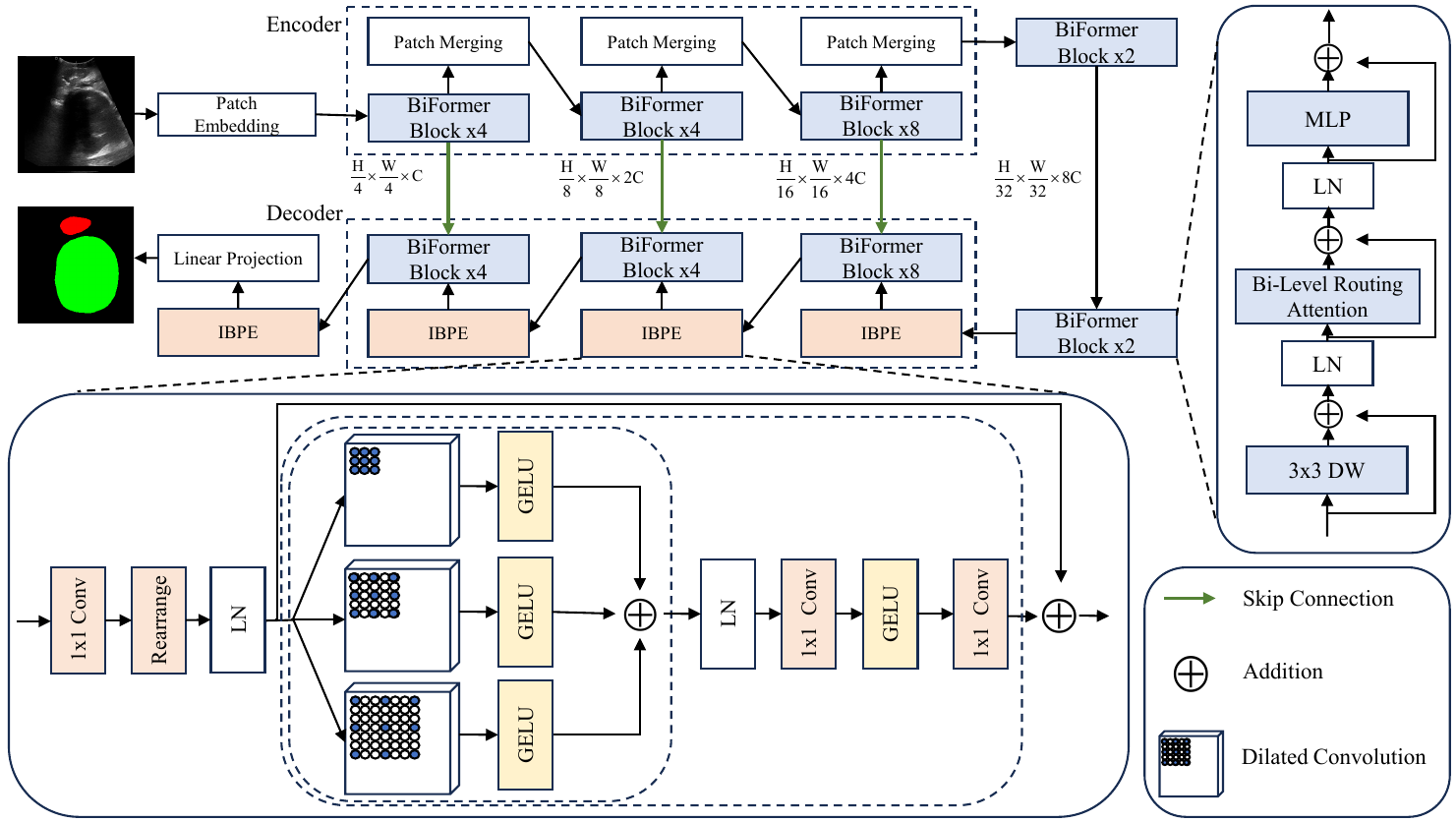}
\caption{The architecture of our proposed BRAU-Net. IBPE, LN and DW represent inverted bottleneck patch expanding, layer normalization and depth-wise convolution, respectively.}
\label{fig2:net}
\end{figure*}

Fig. \ref{fig2:net} shows the network architecture of the proposed BRAU-Net. The network contains three components: encoder, decoder and skip connections. For the encoder, given an  $H \times W \times 3$ input transperineal ultrasound image, it is firstly split into overlapping patches. Subsequently, the token dimension of each patch is projected into $C$ through patch embedding layer. The patch merging layer that is composed of a 3 $\times$ 3 convolutional layer with stride of 2 and padding of 1 and a batch normalization layer, is tasked with decreasing resolution of feature map and increasing dimension. The BiFormer block is tasked with learning feature information. The decoder consists of BiFormer block and inverted bottleneck patch expanding (IBPE) layer. The IBPE module is tasked with reducing the information loss caused by dimensional compression and obtain a better spatial context. Meanwhile, it is also tasked with up-sampling and decreasing dimension. Similar to U-Net \cite{unet}, skip connections are employed to remedy the loss of spatial information caused by down-sampling in the encoder, ensuring the preservation of multiscale features. The last layer of IBPE is used for 4$\times$ up-sampling and replenish of information loss due to dimensional compression, and then a linear projection is utilized to produce segmentation results. More details of our BRAU-Net are elaborated in the following section.
\subsection{BiFormer Block}
BiFormer was proposed in \cite{biformer} that is a novel and general vision transformer. BiFormer block is the core component of BiFormer, which can effectively encode local-global information, as shown in the right box in Fig. \ref{fig2:net}.
The BiFormer block mainly includes a 3 $\times$ 3 depth-wise convolution layer at the beginning, a bi-level routing attention layer at the middle, a 2-layer MLP with expansion ratio $e = 3 $ at the end and 3 residual connections around the three layers. Specifically, the 3 $\times$ 3 depth-wise convolution layer can encode relative position information. The bi-level routing attention is a dynamic, query-aware sparse attention mechanism, which takes into account a small portion of the relevant token in the form of an adaptive query, while avoiding distraction from irrelevant ones. The 2-layer MLP is responsible for modeling relations across positions and embedding for each position, respectively. Residual connection is used to add the output information of previous layer to the output of the current layer.

\subsection{Inverted Bottleneck Patch Expanding Module (IBPE)}
Although the patch expanding layer in SwinUNet \cite{swinunet} is considered more suitable than traditional up-sampling methods (such as nearest-neighbor, bicubic interpolation or deconvolution), it still has a drawback that can not reduce the information loss caused by dimensional compression when transforming information between different dimensional feature spaces. Inspired by the recent success of the structure of inverted bottleneck \cite{mobilenetv2} using small-large-small dimensions in addressing information loss, we also consider adopting this similar idea to solve the problem of information loss encountered during upsampling. Moreover, considering that fine-grained information processing is essential for medical images, we adopt a multi-path dilated convolutional module to increase the receptive fields to obtain a better spatial context. To this end, we design an inverted bottleneck patch expanding module, named IBPE, to replace the patch expanding layer. The following ablation study also prove its necessity for our network.

As illustrated in the down box in Fig. \ref{fig2:net}. In the IBPE module, there are three 1 $\times$ 1 convolutions, a multi-path dilated convolution and a rearrange operation. Taking as an example the second IBPE module in Fig. \ref{fig2:net}, we perform a 1 $\times$ 1 convolutional layer on the input feature maps ($\frac{H}{{16}} \times \frac{W}{{16}} \times 4C$). This operation is used to augment the feature dimension, propelling it to 2$\times$ the input dimension ($4C \to 8C$). Subsequently, a rearrange operation is employed to magnify the resolution of the input feature maps to 2$\times$ the initial resolution while reducing the feature dimension to a mere quarter of the input dimension ($\frac{H}{{16}} \times \frac{W}{{16}} \times 8C$ $\to$ $\frac{H}{{8}} \times \frac{W}{{8}} \times 2C$). Then, the multi-path dilated convolutional module is used to increase the receptive fields to obtain a better spatial context. The number of channels in each path remain unchanged, and the point-wise addition is performed on output channel of each path, and this resulting output ($\frac{H}{{8}}\times\frac{W}{{8}}\times2C$) is used as that of the multi-path expanding convolutional module. Finally, we use two 1 $\times$ 1 convolution to construct an inverted bottleneck structure, whose channels are $8C$ and $2C$, respectively. We will discuss the effects of the IBPE module in the ablation study section.

\subsection{Loss Function}
Throughout the overall training process, a hybrid loss is utilized, merging both dice loss and cross-entropy loss with a balance parameter $\lambda$. The definitions of dice loss ($\cal L$${_{dice}}$), cross-entropy loss ($\cal L$${_{ce}}$), and the hybrid loss($\cal L$) are outlined as follows:
 \begin{equation}
 {\mathcal{L}_{dice}} = 1 - \sum\limits_k^K {\frac{{2{\omega _k}\sum\nolimits_i^N {p(k,i)g(k,i)} }}{{\sum\nolimits_i^N {{p^2}(k,i) + \sum\nolimits_i^N {{g^2}(k,i)} } }}},
 \end{equation}
\begin{equation}
\begin{aligned}
\mathcal{L}_{ce} =-\frac{1}{N}\sum_{i=1}^{N} g(k,i) \cdot \log(p(k,i)) + (1 - g(k,i)) \cdot \log(1 - p(k,i)),
\end{aligned}
\end{equation}
\begin{equation}
\mathcal{L} = \lambda {\mathcal{L}_{dice}} + (1-\lambda) {\mathcal{L}_{ce}},
\end{equation}
where $N$ stands for the number of pixels, where $g(k,i) \in (0,1)$ and $p(k,i) \in (0,1)$ represent the mask and the produced probability for class $k$, respectively. $K$ is the total number of classes, and $\sum\nolimits_k {{\omega _k}}$ = 1 signifies the weight sum of all classes. $\lambda$ serves as a weighted factor, harmonizing the impact of $\mathcal{L}_{dice}$ and $\mathcal{L}_{ce}$. In our work, the parameters ${\omega _k}$ and $\lambda$ are pragmatically set as $\frac{1}{K}$ and 0.6, respectively.

\section{Experiments}
\subsection{Datasets}
To verify the validity of our method, we evaluated the proposed model on two transperineal ultrasound image datasets.

FH-PS-AoP dataset \cite{jnu-ifm}. This dataset is derived from the Pubic Symphysis-Fetal Head Segmentation from Transperineal Ultrasound Images Challenge. The images are 2D B mode ultrasound images and gathered from various locations across China using different ultrasound machines. These images were obtained from pregnant women between 18 and 46 years old. Manual segmentations were conducted using the Pair software, involving a total of 7 annotators, including 2 advisors and 5 graduate and undergraduate students.
The dataset is divided into training and testing sets, consisting of 4000 and 1101 samples in our study, respectively.

HC18 dataset \cite{HC18}. The dataset stands out as the only publicly available dataset with masks specifically dedicated to the FH, consisting of 1354 ultrasound images from 551 pregnant women. 
In the dataset, 999 images are designated for training, while the remaining 355 images are set aside for testing. In contrast to the FH-PS-AOP dataset, the HC18 dataset incorporates FH regions but excludes PS regions. Moreover, since the mask contains only the boundaries of FH, we fill the inside of FH as the new mask of FH.

Notably, since the testing set of HC18 dataset lacks masks, we split the training images in a 9:1 ratio for both training and testing purposes. Also, we adopt a variant of our model , in the PSFHS challenge and win the seventh place on the segmentation task.

\subsection{Evaluation Metrics}
To evaluate the segmentation performance of the proposed BRAU-Net, we utilize three common evaluation metrics: average Dice-Similarity Coefficient (DSC),  average 95$\%$ Hausdorff Distance (HD) and Average Surface Distance (ASD). The DSC is calculated as:
\begin{equation}
\operatorname{DSC} = \frac{{2 \times TP}}{{2 \times TP + FP + FN}},
\end{equation}
where $TP$, $FP$, $FN$ denote true positive, false positive and false negative, respectively. The HD and ASD are formulated as:
\begin{equation}
\operatorname{HD}(Y,\hat Y) = \max \{ \mathop {\max }\limits_{y \in Y} \mathop {\min }\limits_{\hat y \in \hat Y} d(y,\hat y),\mathop {\max }\limits_{\hat y \in \hat Y} \mathop {\min }\limits_{y \in Y} d(y,\hat y)\},
\end{equation}
\begin{equation}
\begin{aligned}
\operatorname{ASD}(Y,\hat Y) = \frac{1}{{|S(Y)| + |S(\hat Y)|}}(\sum\limits_{{s_Y} \in S(Y)} {\mathop {{s_{\hat Y}} \in S(\hat Y)}\limits^{\min } }{d({s_Y},S(\hat Y))} \\ + \sum\limits_{{s_{\hat Y}} \in S(\hat Y)} {\mathop {{s_Y} \in S(Y)}\limits^{\min } d({s_{\hat Y}},S(Y))}),
\end{aligned}
\end{equation}
where $Y$ and $\hat Y$ are the ground truth and predicted segmentation map, respectively. $S(Y)$ and  $S(\hat Y)$ indicate the set of surface voxels of $Y$ and $\hat{Y}$, respectively. $d(\cdot)$ denotes the Euclidean distance.
\subsection{Implementation Details}
The proposed BRAU-Net is implemented on Python 3.7 and PyTorch 1.13.1. The experiments are conducted on an NVIDIA GeForce RTX3060Ti GPU with 8GB memory. During training, we resize the images to 256 $\times$ 256 pixels for the FH-PS-AoP and HC18 datasets, and the data augmentation methods like flipping and rotation are used to increase data diversity. We set batch size as 4 and an initial learning rate as 1e-3 for 150 epochs. The number of attention heads and topks in each stage are 2, 4, 8, 16, 8, 4, 2 and 4, 8, 16, -2, 16, 8, 4, respectively. The Adam optimizer is used to optimize the model for back propagation. We initialize our model with weights pre-trained on ImageNet. In the inference phase, the images of the both datasets are resized to 256 $\times$ 256 pixels.

\subsection{Comparison with Other Methods}
More accurate pubic symphysis-fetal head segmentation can lead to more accurate assessment on the fetal descent process, which is thereby beneficial to reduce the delivery risk of pregnant women. To demonstrate the superiority of the proposed BRAU-Net, we compare it with two CNN-based methods (U-Net \cite{unet} and Attention-UNet \cite{attentionunet}), two methods based on hybrid CNN-Transformer (TransUNet \cite{transunet}, TransAttUNet \cite{transattunet}), and a method based on static sparse Transformer (SwinUNet \cite{swinunet}). The experimental results on the FH-PS-AoP and HC18 datasets are presented in Table \ref{tab1:result}. As can be seen from it, our approach achieves better segmentation performance on DSC and HD evaluation metrics.

\begin{figure*}[htbp]
\centering
\includegraphics[width=1.0\linewidth, keepaspectratio]{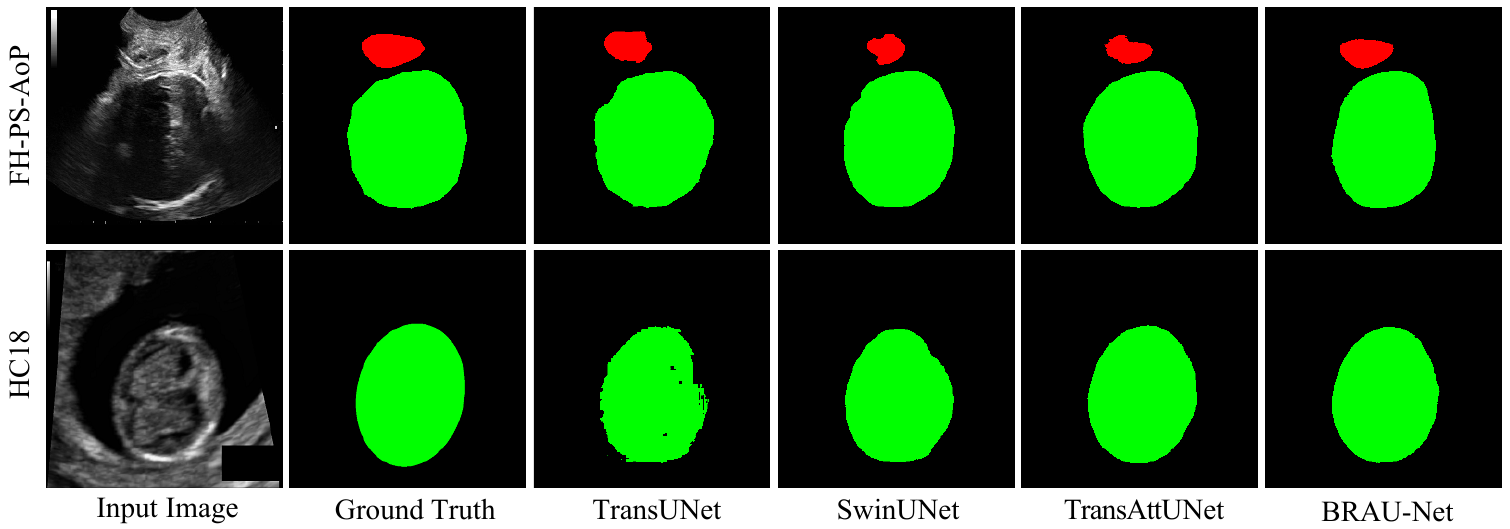}
\caption{The visual segmentation results of our method and others on the FH-PS-AoP and HC18 datasets. The results of our method are closer to ground truth.}
\label{fig3:result}
\end{figure*}

\begin{figure*}[htbp]
\centering
\includegraphics[width=0.5\linewidth, keepaspectratio]{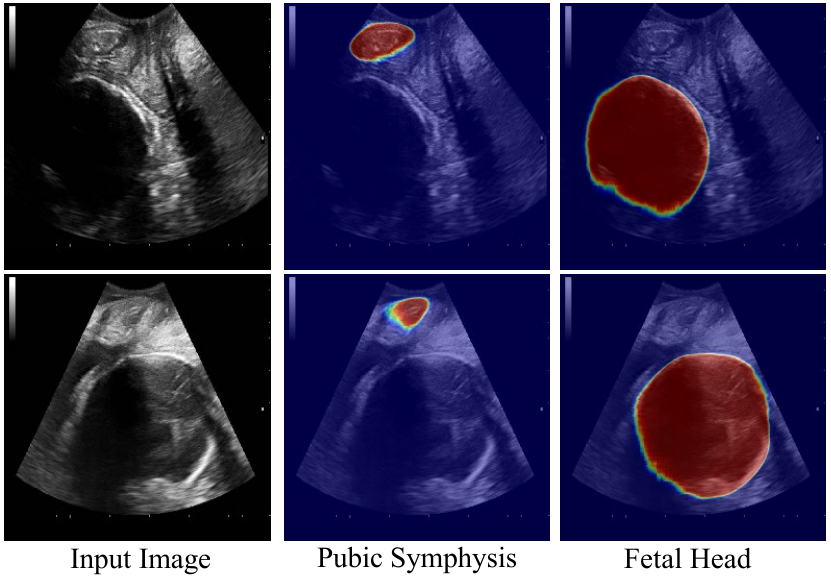}
\caption{The visualization of back-propagated gradient activation maps using Grad-CAM for the final convolution layer on the PS-FH-AoP dataset.}
\label{fig4:relitu}
\end{figure*}

Specifically, compared with U-Net and Attention-UNet on DSC and HD, our method increases 9.98$\%$ and 4.65$\%$ on FH-PS-AoP dataset and decreases 2.4mm and 1.3mm on HC18 dataset, respectively. This shows that transformer-based models have an advantage when it comes to long-range dependency modeling. Compared with TranUNet, TransAttUNet and SwinUNet, our approach improves 2.30$\%$, 1.22$\%$ and 1.77$\%$ in terms of DSC on FH-PS-AoP dataset and reduces 13.77mm, 0.39mm and 1.84mm in terms of HD on HC18 dataset, respectively. It indicates that the dynamic, query-aware sparse attention has more advantages in learning semantic information than vanilla attention and static sparse attention. Meanwhile, the BRAU-Net successfully achieves ASD of 3.70mm and 4.21mm on the both datasets, respectively. The qualitative results are shown in Fig. \ref{fig3:result}, from which it can be seen that our method results in smoother segmentation predictions.
\begin{table}[htbp]
\centering
\caption{The experiment results on the FH-PS-AoP and HC18 datasets. Only DSC is exclusively used for the evaluation of the FH and PS. The symbol $\uparrow$ means the larger the better. The symbol $\downarrow$ means the smaller the better. The best result is in \textbf{Blod}.}
\resizebox{1\columnwidth}{!}{
\begin{tabular}{c|c|c|ccccc|ccc}
\hline
\multirow{2}{*}{Methods} & \multirow{2}{*}{Params (M)} & \multirow{2}{*}{FLOPs (G)}& \multicolumn{5}{c|}{FH-PS-AoP} & \multicolumn{3}{c}{HC18} \\ \cline{4-11} 
                         & &      & DSC $\uparrow$  & HD $\downarrow$ & \multicolumn{1}{l|}{ASD  $\downarrow$} & FH & PS & DSC $\uparrow$     & HD $\downarrow$      & ASD $\downarrow$  \\ \hline
U-Net \cite{unet} & 31.03 & 54.74& 80.22 &14.53 & \multicolumn{1}{l|}{8.23} & 81.42 & 79.02 & 94.14    & 10.52   & 4.73    \\
Attention-UNet \cite{attentionunet}    & 34.88 & 66.63 &  85.55   &12.86   & \multicolumn{1}{l|}{4.37}   &  85.86&   85.24&   \textbf{94.26}   &   9.42      &  4.59    \\
TransUNet \cite{transunet}        & 105.28 &32.30   &  87.90   & 12.24   & \multicolumn{1}{l|}{3.68}    &  88.90  &  86.90  &  90.62      &    21.89     &   8.54    \\
SwinUNet \cite{swinunet}    &  27.17 &  7.73  &  88.43   & 12.13  & \multicolumn{1}{l|}{\textbf{3.42}}    &  89.21  &  87.65  &    93.58    &     9.96    &   4.87    \\
TransAttUNet \cite{transattunet}    &  22.65 &  88.80 &  88.98   & 11.57  & \multicolumn{1}{l|}{3.96}    &  90.36  &  87.60  &    94.04    &     8.51    &   4.68    \\
BRAU-Net &  38.78 & 36.71&  \textbf{90.20}   &  \textbf{10.81}  & \multicolumn{1}{l|}{3.70}  & \textbf{91.04}   & \textbf{89.36} & 94.18&      \textbf{8.12}   &   \textbf{4.21}   \\ \hline
\end{tabular}}
\label{tab1:result}
\end{table}
Additionally, we use Grad-CAM \cite{grad-cam} to generate a \textit{visual justification} for the accurate segmentation decision made by our proposed model, which is an important area of the image displayed from the perspective of a specific layer using back-propagated gradient. In Fig. \ref{fig4:relitu}, we visualize the heat map of the class activation map for the final convolution layer on the PS-FH-AOP dataset. We can see that our model produces good activation signals in the PS and FH regions. This qualitatively gives an explainability of our model. 
\subsection{Ablation Study}
In this section, we explore the effects of patch expanding layer and IBPE on segmentation results. Table \ref{tab2:abalation} shows the results of ablation studies. As presented in Table \ref{tab2:abalation}, compared to the patch expanding layer, the IBPE brings a performance improvement under all evaluation metrics. Specifically, with respective to FH-PS-AoP dataset, the IBPE improves 0.78$\%$ on DSC and reduces 0.87mm on HD and 0.28mm on ASD than patch expanding layer, respectively. With respective to HC18 dataset, the IBPE outperforms patch expanding by 0.73$\%$ on DSC. This shows that the IBPE can reduce information loss and gain better spatial context. From Table \ref{tab1:result} and Table \ref{tab2:abalation}, we can see that the DSC of BRAU-Net with patch expanding layer and IBPE  improves  by 0.99$\%$ and 1.77$\%$ against that of  SwinUNet on FH-PS-AoP dataset, respectively.
This suggests that our sparse transfomer and IBPE module are better suitable for restoring resolution and increasing dimension to feature maps than patch expanding layer for medical image segmentation tasks.
\begin{table*}[htbp]
\centering
\caption{The results of ablation study on the FH-PS-AoP and HC18 datasets.}
\resizebox{1\columnwidth}{!}{
\begin{tabular}{c|c|c|ccccc|ccc}
\hline
\multirow{2}{*}{Up-sampling} & \multirow{2}{*}{Params (M)} & \multirow{2}{*}{FLOPs (G)}& \multicolumn{5}{c|}{FH-PS-AoP}                & \multicolumn{3}{c}{HC18} \\ \cline{4-11} 
                         &   &  & DSC $\uparrow$  & HD $\downarrow$ & \multicolumn{1}{l|}{ASD  $\downarrow$} & FH & PS & DSC $\uparrow$     & HD $\downarrow$      & ASD $\downarrow$  \\ \hline
Patch expanding    & 33.30& 14.23  & 89.61 &11.42   & \multicolumn{1}{l|}{3.94}    & 90.24&    88.98&  93.45  &   8.64      &   4.57   \\
IBPE       & 38.78 & 36.71  &  \textbf{90.20}  & \textbf{10.81}   & \multicolumn{1}{l|}{\textbf{3.70}}    &  \textbf{91.04}  &  \textbf{89.36}  & \textbf{94.18}      &    \textbf{8.12}    &   \textbf{4.21}     \\ \hline
\end{tabular}}
\label{tab2:abalation}
\end{table*}

\subsection{Analysis on Computational Complexity}
\begin{figure*}[htbp]
\centering
\includegraphics[width=0.8\linewidth, keepaspectratio]{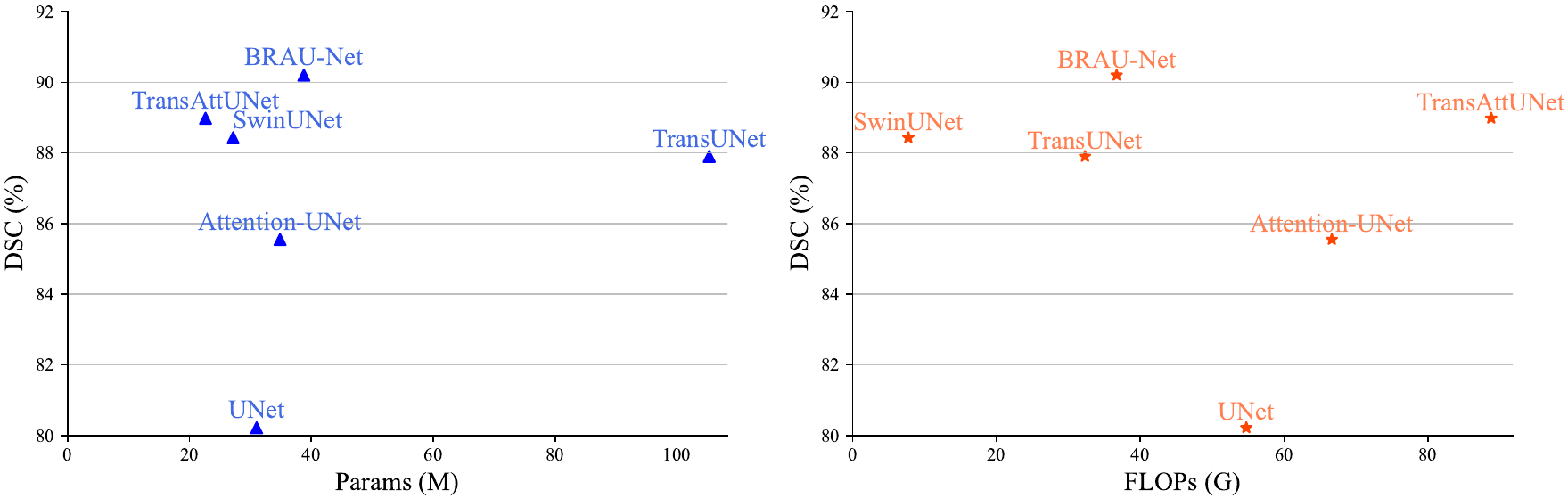}
\caption{Comparisons of efficiency vs performance of BRAU-Net against other methods.}
\label{fig5:param}
\end{figure*}

To further demonstrate the validity of our model, we compare our model with others in terms of the number of parameters and floating point operations (FLOPs). The FLOPs are calculated with the image of resolution 256 $\times$ 256 on FH-PS-AoP dataset. These results are shown in Table \ref{tab1:result}. One can see from Table \ref{tab1:result} that with respective to Params and FLOPs, although our method has a relatively higher (but not a highest) value than other ones, our method achieves a best performance. 
In order to further intuitively compare our model with others on the efficiency and performance, we take the relationships between DSC and parameters, as well as DSC and FLOPs as example to illustrate, as shown in Fig. \ref{fig5:param}. The adopted DSC is also evaluated on the  FH-PS-AoP dataset. As can be seen from Fig. \ref{fig5:param} that our method has a comparable number of parameters compared to Attention-UNet \cite{attentionunet}, similarly, we have also comparable FLOPs with TransUNet \cite{transunet} (36.71G vs. 32.30G), but our method all achieves a higher DSC, which shows that our method is relatively more efficient.
\subsection{Analysis on Network Depth and Width}
As two hyperparameter of BRAU-Net, network depth and width are directly related to the number of parameters, FLOPs and segmentation performance. We explore the effect of four different depth or width on segmentation performance on FH-PS-AoP dataset. The results are shown in Table \ref{tab:deep&width}. Specifically, it can be observed that compared to BRAU-Net-S, although BRAU-Net-T reduces the number of channels of the network, that is, the width of the network, thereby resulting in fewest parameters (8.46M) and FLOPs (18.97G), it also achieves fewest segmentation results accordingly. Additionally, compared to BRAU-Net-S, we increase the depth of BRAU-Net network, correspondingly the DSC increases by 1.16$\%$ and the HD decreases by 1.15mm. Also, we increase the number of blocks in the 3-$rd$ and 5-$th$ stages, resulting in a higher performance, which indicates that appropriate the number of blocks in depth can better capture advanced semantic information.
\begin{table*}[htbp]
\centering
\caption{The results of depth and width of different variants.}
\resizebox{1\columnwidth}{!}{
\begin{tabular}{c|c|c|c|c|ccc|cc}
\hline
Method &Channels  & Blocks & Params (M) & FLOPs (G) & DSC $\uparrow$ & HD $\downarrow$ & ASD $\downarrow$ & FH  & PS \\ \hline
  BRAU-Net-T     & 64 &  [2,2,2,2,2,2,2]&  8.46&  12.93&  88.20& 12.04 & 4.05 & 88.62 &87.78  \\
  BRAU-Net-S     & 96 &[2,2,2,2,2,2,2]& 18.97  & 28.97 & 89.04 & 11.96 &  3.89& 89.56 & 88.52 \\
  BRAU-Net-B     &96  &[4,4,4,4,4,4,4]  & 26.84 & 33.65 & 89.57 & 11.51 & \textbf{3.69} & 90.45 &88.69  \\ 
  BRAU-Net     &96  &[4,4,8,4,8,4,4]  & 38.78 & 36.71 & \textbf{90.20} & \textbf{10.81} & 3.70 & \textbf{91.04}&\textbf{89.36}  \\ \hline
\end{tabular}}
\label{tab:deep&width}
\end{table*}

\subsection{Analysis on Pre-trained Weights}
In the PSFHS challenge, we apply BRAU-Net without pre-training weights initialization and a patch expanding layer as up-sampling to train model. In this section, we explore the effect of pre-trained weights on segmentation performance. The results are shown in Table \ref{tab:pre}. It can be seen that the model with pre-training can achieve a better segmentation than the model without pre-trained under all the metrics. Meanwhile, the BRAU-Net with pre-training can better identify FH target. This qualitatively result shows the necessity of pre-training model to improve the performance of ultrasonic image segmentation.
\begin{table}[htbp]
\centering
\caption{The results of the impact of pre-trained weights on performance.}
\begin{tabular}{c|ccc|cc}
\hline
pre-trained  & DSC $\uparrow$ & HD $\downarrow$ & ASD $\downarrow$ & FH & PS \\ \hline

   -- & 89.06    &  11.45  & 3.97   &  89.41  & 88.71 \\
    $\checkmark$ & \textbf{89.61}   & \textbf{11.42}  &   \textbf{3.94} &  \textbf{90.24}  &  \textbf{88.98}  \\\hline
\end{tabular}
\label{tab:pre}

\end{table}

\section{Conclusion}
This study proposes a novel network, called BRAU-Net for the pubic symphysis-fetal head segmentation, whose core components are BiFormer block and IBPE module. The former can efficiently model the long-range dependency in a query-aware manner while reducing the computational complexity, the latter can reduce the information loss caused by dimensional compression and obtain a better spatial context. Extensive experimental results show that our method can achieve excellent segmentation performance on the FH-PS-AoP and HC18 datasets. In the future, we will design more powerful dynamic attention mechanism for ultrasonic image segmentation.

\section*{Acknowledgments}
This work is supported in part by the Scientific Research Foundation of Chongqing University of Technology under Grants 0103210650 and 0121230235, and in part by the Youth Project of Science and Technology Research Program of Chongqing Education Commission of China under Grants KJQN202301145 and KJQN202301162. We would like to thank the anonymous reviewers for their helpful comments which have led to many improvements in this paper.

%
%
%
%




\clearpage

\begin{thebibliography}{10}

\bibitem{RN1}
Myra Fitzpatrick, Kathryn McQuillan, and Colm O’Herlihy.
\newblock Influence of persistent occiput posterior position on delivery outcome.
\newblock {\em Obstetrics \& Gynecology}, 98(6):1027--1031, 2001.

\bibitem{RN2}
Penny Simkin.
\newblock The fetal occiput posterior position: state of the science and a new perspective.
\newblock {\em Birth}, 37(1):61--71, 2010.

\bibitem{RN4}
Sherer DM, Bradley KS, and Langer O.
\newblock Intrapartum fetal head position i: comparison between transvaginal digital examination and transabdominal ultrasound assessment during the active stage of labor.
\newblock {\em Ultrasound in Obstetrics and Gynecology}, 19(3):258--263, 2002.

\bibitem{fcn}
Jonathan Long, Evan Shelhamer, and Trevor Darrell.
\newblock Fully convolutional networks for semantic segmentation.
\newblock In {\em Proceedings of the IEEE conference on computer vision and pattern recognition}, pages 3431--3440, 2015.

\bibitem{unet}
Olaf Ronneberger, Philipp Fischer, and Thomas Brox.
\newblock U-net: Convolutional networks for biomedical image segmentation.
\newblock In {\em Medical image computing and computer-assisted intervention--MICCAI 2015: 18th international conference, Munich, Germany, October 5-9, 2015, proceedings, part III 18}, pages 234--241. Springer, 2015.

\bibitem{attentionunet}
Ozan Oktay, Jo~Schlemper, Loic~Le Folgoc, Matthew Lee, Mattias Heinrich, Kazunari Misawa, Kensaku Mori, Steven McDonagh, Nils~Y Hammerla, Bernhard Kainz, et~al.
\newblock Attention u-net: Learning where to look for the pancreas.
\newblock {\em arXiv preprint arXiv:1804.03999}, 2018.

\bibitem{muititask}
Yaosheng Lu, Dengjiang Zhi, Minghong Zhou, Fan Lai, Gaowen Chen, Zhanhong Ou, Rongdan Zeng, Shun Long, Ruiyu Qiu, Mengqiang Zhou, et~al.
\newblock Multitask deep neural network for the fully automatic measurement of the angle of progression.
\newblock {\em Computational and mathematical methods in medicine}, 2022(1):5192338, 2022.

\bibitem{DBSN}
Jieyun Bai, Zhanhang Sun, Sheng Yu, Yaosheng Lu, Shun Long, Huijin Wang, Ruiyu Qiu, Zhanhong Ou, Minghong Zhou, Dengjiang Zhi, et~al.
\newblock A framework for computing angle of progression from transperineal ultrasound images for evaluating fetal head descent using a novel double branch network.
\newblock {\em Frontiers in physiology}, 13:940150, 2022.

\bibitem{transformer}
Ashish Vaswani, Noam Shazeer, Niki Parmar, Jakob Uszkoreit, Llion Jones, Aidan~N Gomez, {\L}ukasz Kaiser, and Illia Polosukhin.
\newblock Attention is all you need.
\newblock {\em Advances in neural information processing systems}, 30, 2017.

\bibitem{transunet}
Jieneng Chen, Yongyi Lu, Qihang Yu, Xiangde Luo, Ehsan Adeli, Yan Wang, Le~Lu, Alan~L Yuille, and Yuyin Zhou.
\newblock Transunet: Transformers make strong encoders for medical image segmentation.
\newblock {\em arXiv preprint arXiv:2102.04306}, 2021.

\bibitem{transfuse}
Yundong Zhang, Huiye Liu, and Qiang Hu.
\newblock Transfuse: Fusing transformers and cnns for medical image segmentation.
\newblock In {\em Medical image computing and computer assisted intervention--MICCAI 2021: 24th international conference, Strasbourg, France, September 27--October 1, 2021, proceedings, Part I 24}, pages 14--24. Springer, 2021.

\bibitem{hiformer}
Moein Heidari, Amirhossein Kazerouni, Milad Soltany, Reza Azad, Ehsan~Khodapanah Aghdam, Julien Cohen-Adad, and Dorit Merhof.
\newblock Hiformer: Hierarchical multi-scale representations using transformers for medical image segmentation.
\newblock In {\em Proceedings of the IEEE/CVF winter conference on applications of computer vision}, pages 6202--6212, 2023.

\bibitem{transformerrmeetunet}
Reza Azad, Moein Heidari, Yuli Wu, and Dorit Merhof.
\newblock Contextual attention network: Transformer meets u-net.
\newblock In {\em International Workshop on Machine Learning in Medical Imaging}, pages 377--386. Springer, 2022.

\bibitem{swinunet}
Hu~Cao, Yueyue Wang, Joy Chen, Dongsheng Jiang, Xiaopeng Zhang, Qi~Tian, and Manning Wang.
\newblock Swin-unet: Unet-like pure transformer for medical image segmentation.
\newblock In {\em European conference on computer vision}, pages 205--218. Springer, 2022.

\bibitem{swintransformer}
Ze~Liu, Yutong Lin, Yue Cao, Han Hu, Yixuan Wei, Zheng Zhang, Stephen Lin, and Baining Guo.
\newblock Swin transformer: Hierarchical vision transformer using shifted windows.
\newblock In {\em Proceedings of the IEEE/CVF international conference on computer vision}, pages 10012--10022, 2021.

\bibitem{medt}
Jeya Maria~Jose Valanarasu, Poojan Oza, Ilker Hacihaliloglu, and Vishal~M Patel.
\newblock Medical transformer: Gated axial-attention for medical image segmentation.
\newblock In {\em Medical image computing and computer assisted intervention--MICCAI 2021: 24th international conference, Strasbourg, France, September 27--October 1, 2021, proceedings, part I 24}, pages 36--46. Springer, 2021.

\bibitem{biformer}
Lei Zhu, Xinjiang Wang, Zhanghan Ke, Wayne Zhang, and Rynson~WH Lau.
\newblock Biformer: Vision transformer with bi-level routing attention.
\newblock In {\em Proceedings of the IEEE/CVF conference on computer vision and pattern recognition}, pages 10323--10333, 2023.

\bibitem{nunet}
Gongping Chen, Lei Li, Jianxun Zhang, and Yu~Dai.
\newblock Rethinking the unpretentious u-net for medical ultrasound image segmentation.
\newblock {\em Pattern Recognition}, 142:109728, 2023.

\bibitem{smunet}
Zhenyuan Ning, Shengzhou Zhong, Qianjin Feng, Wufan Chen, and Yu~Zhang.
\newblock {SMU-Net}: Saliency-guided morphology-aware u-net for breast lesion segmentation in ultrasound image.
\newblock {\em IEEE transactions on medical imaging}, 41(2):476--490, 2021.

\bibitem{fh-pssnet}
Zhensen Chen, Zhanhong Ou, Yaosheng Lu, and Jieyun Bai.
\newblock Direction-guided and multi-scale feature screening for fetal head--pubic symphysis segmentation and angle of progression calculation.
\newblock {\em Expert Systems with Applications}, 245:123096, 2024.

\bibitem{mobilenetv2}
Mark Sandler, Andrew Howard, Menglong Zhu, Andrey Zhmoginov, and Liang-Chieh Chen.
\newblock Mobilenetv2: Inverted residuals and linear bottlenecks.
\newblock In {\em Proceedings of the IEEE conference on computer vision and pattern recognition}, pages 4510--4520, 2018.

\bibitem{jnu-ifm}
Yaosheng Lu, Mengqiang Zhou, Dengjiang Zhi, Minghong Zhou, Xiaosong Jiang, Ruiyu Qiu, Zhanhong Ou, Huijin Wang, Di~Qiu, Mei Zhong, et~al.
\newblock The jnu-ifm dataset for segmenting pubic symphysis-fetal head.
\newblock {\em Data in brief}, 41:107904, 2022.

\bibitem{HC18}
Thomas~LA van~den Heuvel, Dagmar de~Bruijn, Chris~L de~Korte, and Bram~van Ginneken.
\newblock Automated measurement of fetal head circumference using 2d ultrasound images.
\newblock {\em PLoS One}, 13(8):e0200412, 2018.

\bibitem{transattunet}
Bingzhi Chen, Yishu Liu, Zheng Zhang, Guangming Lu, and Adams Wai~Kin Kong.
\newblock {TransAttUnet}: Multi-level attention-guided u-net with transformer for medical image segmentation.
\newblock {\em IEEE Transactions on Emerging Topics in Computational Intelligence}, 2023.

\bibitem{grad-cam}
Ramprasaath~R Selvaraju, Michael Cogswell, Abhishek Das, Ramakrishna Vedantam, Devi Parikh, and Dhruv Batra.
\newblock Grad-{CAM}: Visual explanations from deep networks via gradient-based localization.
\newblock In {\em Proceedings of the IEEE international conference on computer vision}, pages 618--626, 2017.

\end{thebibliography}

\end{document}